# WIKIPULSE – A NEWS-PORTAL BASED ON WIKIPEDIA

Tobias Futterer, Karsten Packmohr, Stefan Schultheiss
University of Bamberg
Bamberg, Germany
futterer.tobias@gmail.com
karsten.packmohr@googlemail.com, stefan-reinhold.schultheiss@stud.uni-bamberg.de

Tushar Malhotra, Harrison Mfula
Aalto University
Espoo, Finland
harrison.mfula@aalto.fi
tushar.malhotra@aalto.fi

Peter A. Gloor
MIT CCI
5 Cambridge Center
Cambridge MA 02138
pgloor@mit.edu

## ABSTRACT

More and more user-generated content is complementing conventional journalism. While we don't think that CNN or New York Times and its professional journalists will disappear anytime soon, formidable competition is emerging through humble Wikipedia editors. In earlier work (Becker 2012), we found that entertainment and sports news appeared on average about two hours earlier on Wikipedia than on CNN and Reuters online. In this project we build a news-reader that automatically identifies late-breaking news among the most recent Wikipedia articles and then displays it on a dedicated Web site.

## 1. INTRODUCTION

Wikipedia, one of the most important web 2.0 websites is available in more than 280 different languages and contains over 22 million articles with about 18.4 million registered users, 77,000 are active contributors that collectively work on Wikipedia. Wikipedia contributors are spread all over the world, together they create a 24/7 online community. This community quickly creates articles based on news coming from various news sources, with some articles even written by Wikipedians involved into the actual events (Iba et al. 2009).

Earlier studies found that Wikipedia is in some cases faster than conventional news channels (Becker, 2012). These observations formed the foundation of the Wikipulse project and prompted Gloor et al. (2012) to propose the use of Wikipedia content to find "latest trends based on the analysis of recent edits on Wikipedia articles." Wikipulse aims at complementing other news sources by generating latest news based on Wikipedia article edits and presenting them in a user friendly news format.

The main contributions of this paper are the Wikipulse algorithm which shows how to use Wikipedia to generate news automatically, and the description of a first implementation.

The remainder of this paper is organized as follows: Section 2 presents related work. Section 3 discusses the architecture; Section 4 introduces the news section algorithm while Section 5 talks about data analysis. Section 6 discusses the current implementation. Finally, Section 7 lays the foundation for future work.

## 2. RELATED WORK

In their research studying the collaborative behavior of Wikipedia editors, Bayer et.al (2011) found that unlike just many eyes having a look at an article, the experience of the editors is important – they should have worked on many other articles for the quality of their articles to be good. It was also found that a high number of editorial events contribute positively to a page's quality. In other earlier work (Becker 2012), it was found that entertainment and sports news appeared on average about two hours earlier on Wikipedia than on CNN and Reuters online. Wikirage, another Wikipedia-based news system, tracks the pages in Wikipedia which are receiving the most edits over various periods of time (Wood 2011). While this site does a good job collecting the edits it does not process the results further and as evidenced in (Bayer et al. 2012) edits alone are not enough to justify newsworthiness. Nevertheless, Wikirage delivers a good benchmark to validate against the results of our news generation algorithm.

## 3. ARCHITECTURE

In order to build a Wikipedia-based news portal, three major tasks need to be addressed:
1. Find relevant articles on Wikipedia

2. Reformat the articles in news style format and
3. Display them on a Web page.

The core of the system consists of the following main parts: Wikipedia acts as the primary source of news items through articles edited and published by the Wikipedia community, secondly, the Wikipulse news generation algorithm automatically finds the most newsworthy articles on Wikipedia grouped by Wikipedia categories (e.g. Current Events, Sports, Politics etc).

Besides obvious selection criteria such as the most recently edited and searched articles, in the first step mentioned above, we also employ algorithms from earlier work (Fuehres 2012), where we discovered that building an article network based on "shared-editorship" links – two articles obtain a link if the same editor edits both of them – points out the most important recent articles. For instance, editing a list of cricket players from 1900 might lead to many edits, but such a list is obviously not newsworthy. Selecting the "right authors" guarantees newsworthy articles by maintaining a continuously growing watchlist of frequent editors who focus on news-type article. Fact-checking of the article happens for free through the many-eyeball principle of Wikipedia. While there are stub-level articles of low quality, heavily edited and accessed articles are continuously checked by the Wikipedian peer group.

In the second step we reformat the Wikipedia articles, which are written in factual history-style, into a more journalistic-style, using automatic abstract generation techniques. The final step consists of displaying the articles in a reader-friendly online newspaper. The figure below illustrates the major system components. From top to bottom, the Web frontend represents the browser that serves as a client interaction point with the system, next, the News Selection and Extraction components are the core processing units of the system. The integration component to Wikipedia is shown at the bottom of figure 1.

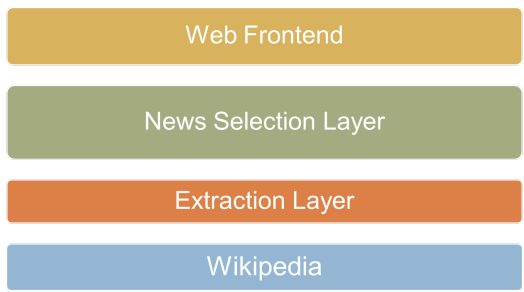

Figure 1: Basic design of Wikipulse.

The Wikipulse application uses a web centric 3 tier layered architecture described in figure 2. Tier 1 also known as the client tier uses the browser to represent the user's point of interaction with the application. In order to promote modularity and manageability, tier 2 is subdivided into logical layers, namely presentation, service, identification, extraction and data access layers. The presentation layer houses the application's parsing and presentation logic. The service layer is a façade design pattern implementation which consolidates different underlying APIs into a uniform service processed by the presentation layer. The Identification layer, as the name suggests, is responsible for news identification and page ranking. It also takes care of the generation of appropriate news summaries/excerpts that accompany each ranked news-item. The extraction layer – a wrapper implementation of the Wikipedia API, is used to communicate with and extract information from Wikipedia. The data access layer encapsulates all access to and from external data storage providing a uniform and database agnostic interface to the data tier. Tier 3 represents the persistent data store needed for storing and retrieving data used by the application.

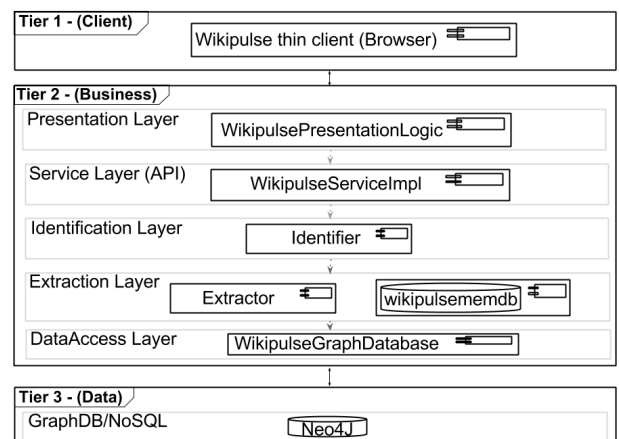

Figure 2: Wikipulse 3-tier Architecture

## 4. NEWS SELECTION ALGORITHM

The news algorithm is responsible for selecting Wikipedia pages and creating the news objects. It consists of multiple steps:
a. working set creation
b. news selection
c. news creation
d. news saving

*a. working set creation*
The algorithm runs periodically and each run processes a set of pages which are called the "working set". This working set contains recently edited pages in Wikipedia. Each page of a working set is passed through and processed by the other parts of the algorithm.

First the page metadata, the id, the title, all of its authors and all of its categories are saved in the authorgraph-database, making it possible for the collaborating parts of the algorithm to execute various queries. Then the "news selection"- part of the algorithm ranks and selects the pages from the working set. These pages are converted to news by the "news creation"-algorithm. Finally the generated news items are saved into the database.
The saving procedure ensures that the database is updated and expanded with each run of the algorithm even if no news items were generated.

*b. news selection*
The news selection process finds Wikipedia pages that become news items later on. It does this by generating multiple ranks $r$ for each page, summarizing them and comparing the result to a threshold value $t$. Each rank is based on database queries and weighted by a specific "rank-weight" $w$ to influence its importance on the final result. A news item is generated from a page when the following condition is met:
$$\sum_{i=1}^{n} r_i * w_i > t_i \quad (1)$$

Each different rank that is generated for each page is described in the sections below. Most of the ranks are ratios computed in order to have a predictable set of numbers as a result. The implementation design of the algorithm is very modular, so it is possible to add further ranking-mechanisms easily.

*AuthorsWithNews*
The AuthorsWithNews-rank increases the importance of pages which have been edited by news-generating authors, that is, authors we already have generated news from.
It calculates the ratio of the number of news-generating authors that edited the current page $p$ to the number of all news-generating authors $a_n$:

$$r_0(p) = \frac{a_n(p)}{a_n} \quad (2)$$

*CommonAuthors*
The CommonAuthors-rank calculates the popularity of a Wikipedia page among the authors. It accomplishes this by computing the ratio of all authors $a$ editing the current page to all authors in the database:

$$r_1(p) = \frac{a(p)}{a} \quad (3)$$

*DomainExperts*
The DomainExperts-rank identifies important pages by looking at the amount of domain experts that edited that page. A domain expert is a Wikipedia author who edits pages with the same categories as the page that is being evaluated. The page is ranked by creating the ratio of all page-authors that have edited other pages with the same category $a_{c(p)}$ to all authors in the database:
$$r_2(p) = \frac{a_{c(p)}(p)}{a} \quad (4)$$

*RecentChanges*
This rank measures the editing activity on the current page in comparison to all other pages in the working set. This is done by calculating the average number of edits in this working set and then comparing the edit count of the current page to it.

*Relevance*
The relevance rank uses the "stats.grok.se"-webservice to determine the popularity of a Wikipedia page. It creates the rate of page views yesterday in relation to page views over the last 30 days.

As described above, all ranks are calculated, weighted and then summarized into a final rank. If this final rank is greater than the threshold the page gets handed to the next part of the news generation algorithm:

*c. news creation*
The news creation process creates news from pages. It does this by analyzing and aggregating the edits of a page into a big text block, which is then sent to the "summry.com"-webservice to be summarized into a short news article. This news article is then, along with some metadata, saved into the database and presented via the web interface when needed.

The edits to be aggregated are selected based on three criteria:
1. The edit was made by a top editor.
2. The edit was made by a domain expert
3. The edit is longer than 50 characters

A top editor is an author who is in the "top 50 edits"-list and domain experts have the most edits for a given category.

**4.1 Extraction**
The Extraction layer provides access to Wikipedia and other supporting external resources. The implemented API among others gives access to Wikipedia's recent changes, page, edit and author information.

The layer also makes use of the following supporting interfaces:
- smmry.com – an open API used for summarizing text of pages exceeding an agreed number of sentences (http://smmry.com/api).
- stats.grok.se – a Wikipedia sub project which maintains page statistics for Wikipedia pages (http://stats.grok.se/json/en/latest30/).

For data access the layer uses interfaces to the Hsql in-memory database for data caching and a data access API to the NoSQL graph database Neo4J.

## 5. DATA ANALYSIS

In order to measure the performance of the news-selection algorithm/logic and improve it iteratively, it was decided to benchmark Wikipulse against some other, traditionally well-recognized news sources (N) (e.g. CNN, BBC, Reuters, AP etc.). The general idea was to analyze whether the news-items 'selected' and 'ranked' by Wikipulse were also being reported as news by the conventional media (within admissible time bounds). In particular, the goal is to benchmark on the following abstract criteria:

### Accuracy and Relevance

This is a measure of 'overlap' between the news reported by Wikipulse and the one reported by the chosen traditional news source (N) at any given time. The chosen news source (N) is assumed to be the gold standard for relevance and accuracy and hence a high degree of overlap is considered to indicate better performance of the Wikipulse algorithm.

### Freshness/Speed

This criteria measures the relative temporal ordering of the overlapping news-items between Wikipulse and the chosen conventional news source (N). This indicates whether Wikipulse is faster or slower than the chosen benchmark in terms of reporting.

### Implementation

A standalone benchmarking tool was developed to compare and report performance on the above mentioned criteria. The tool uses the RSS feeds published by the various traditional news media/sources to compare the news feed (custom, non-RSS) generated by Wikipulse.

### Calculating Overlap - Identifying 'matching' news stories

The key part of the benchmarking logic deals with calculating the extent of overlap between the Wikipulse news feed and that of other candidate news feeds. In the current implementation, this overlap is defined as the percentage of matching news items in the two candidate news feeds (Wikipulse and another feed). However, locating matching news items or stories in two different news feeds is non-trivial because the matches need to be determined at a semantic level and not textually. Two stories discussing or reporting the same news can be worded completely different from each other.

In order to overcome this obstacle, for the purpose of the proof-of-concept implementation, a keyword based matching heuristic was adopted which works as follows: Individual stories from both the feeds are processed to obtain a list of keywords. At present, the keyword extraction is done using the publicly available smmry.com service (already discussed earlier in the document). The news items/stories in both the feeds are then compared for matching keywords. A 'Match' is expressed in terms of 'match strength' - a fraction between 0 and 1 with 0 indicating no match and 1 indicating a perfect or exact match. The match strength is the number of matching keywords between two given stories to the total number of keywords in either story.

After analyzing a number of test runs between various pairs of news feeds (e.g. BBC and Reuters, Reuters and CNN, CNN and AP etc.), a minimum match strength threshold was established. News items from the two candidate feeds with a match strength greater than the established threshold are classified as 'matches' and count towards calculating the 'overlap'. Based on the tests, 0.25 was established as the minimum threshold to avoid missing any matches. However, matches with strengths between 0.25 and 0.33 also resulted in a lot of false positives. Match strengths greater than 0.33 generally indicated a strong match.

The task of locating matching news stories between feeds and calculating the overlap is non-trivial. While the current heuristics, described above, perform reasonably well, the tool still causes both **missed matches** (i.e. matching stories which are not reported by the tool) as well as **false positives** (i.e. unrelated stories reported as matches by the tool). Besides, there are other challenges involved with processing the feeds. E.g. in a given news feed, there may be multiple smaller stories related to the same broader theme or topic. Ideally, these individual but related news-items should be considered as a single unit for the purpose of comparison. This would require advanced semantic analysis of the news-items.

Hence, the future versions of the benchmarking tool should use more sophisticated heuristics, drawing from NLP and semantic analysis, to yield more precise results.

## 6. DISCUSSION

From current implementation and tests, the results show that it is possible to generate latest news based on Wikipedia articles. The results also show that there are various criteria that can be used to classify article edits used as news sources. Frequency of edit counts can lead to higher quality of articles and experience of editors plays a major role in determining article quality. News items produced by Wikipulse can vary depending on the filtering rules applied and other related factors. Due to the complexity of the relationships between the editors and articles, it is possible to have some news items generated which are completely irrelevant. Some results obtained in this research might certainly benefit from further verification and fine tuning.

## 7. FUTURE WORK

Going forward, the work done during this project can be extended in many ways. For example we might embed the benchmarking module as part of the self-learning and correction process of the algorithm. We also consider employing known open source text processing libraries like lucene (http://lucene.apache.org/core/) and Stanford NLP (http://nlp.stanford.edu /software/index.shtml) for text manipulation. These libraries were left out in this phase of the project in order to reduce complexity.

The news-selection and news-creation algorithms need further long-term tests so parameters can be tuned to improve the result. The news-algorithm can also be enhanced by integrating user interaction and manual intervention.

Currently the system is based on the categories provided by Wikipedia. Since the category structure is relatively wide and often not useful, a future task would be to create a specific set of categories for Wikipulse and merge it with the Wikipedia category structure to assign news to more general and appropriate categories during the news-creation process.

Right now the news-selection is partly based on the network structure of the author and his/her amount of news. Wöhner et al. (2011) propose to use the previous behavior of participants to measure and estimate their "newsworthiness", e.g. what they usually change (fixing mistakes, adding content, adding references etc.). Suzuki and Yoshikawa (2012) also propose an author evaluation.

The news-creation process is currently partly outsourced to smmry.com but there is no control of the news quality. It is therefore necessary to use additional factors in the future to measure and validate the quality. Blumenstock (2008) lists word count as one possible way, while Zeng et.al (2006) propose to use the edit history of an article to measure possible outcome.

The findings and experiences presented in this paper open an exciting window to future work and more improvement opportunities for automatic Wikipedia-based news generation.

**APPENDIX – SCREEN SHOTS OF WIKIPULSE PROTOTYPE**

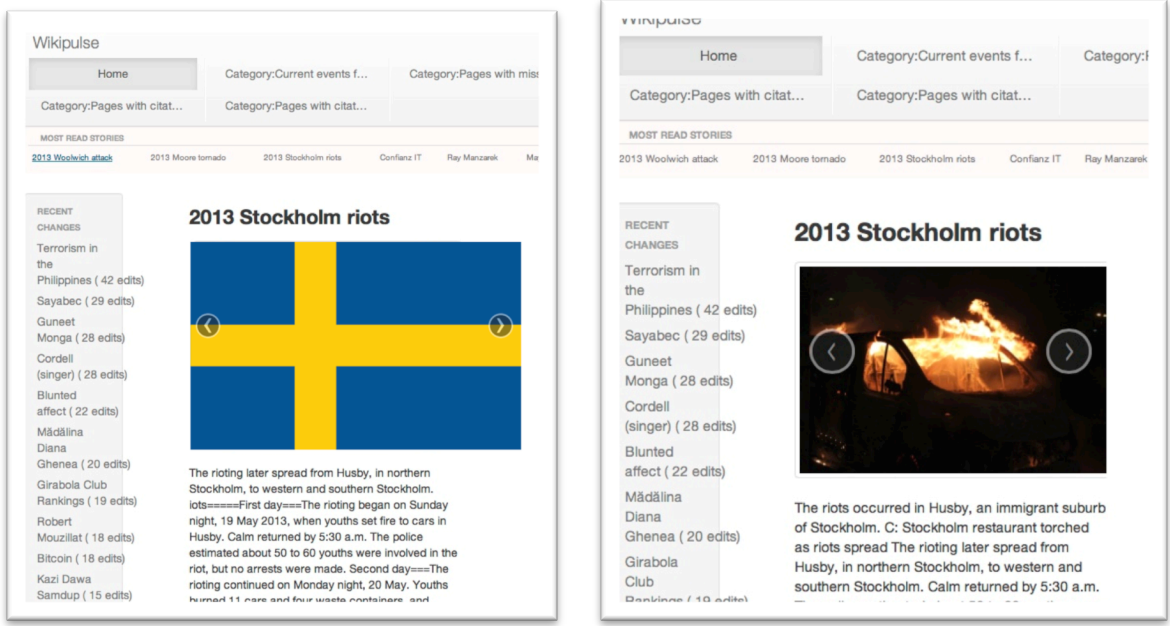

Figure 3. 2 Screen shots of the Wikipulse entry on the Stockholm riots starting May 19 to 28 auto-generated on May 24, 2012 at 8PM (right) and May 25, 9AM (left)

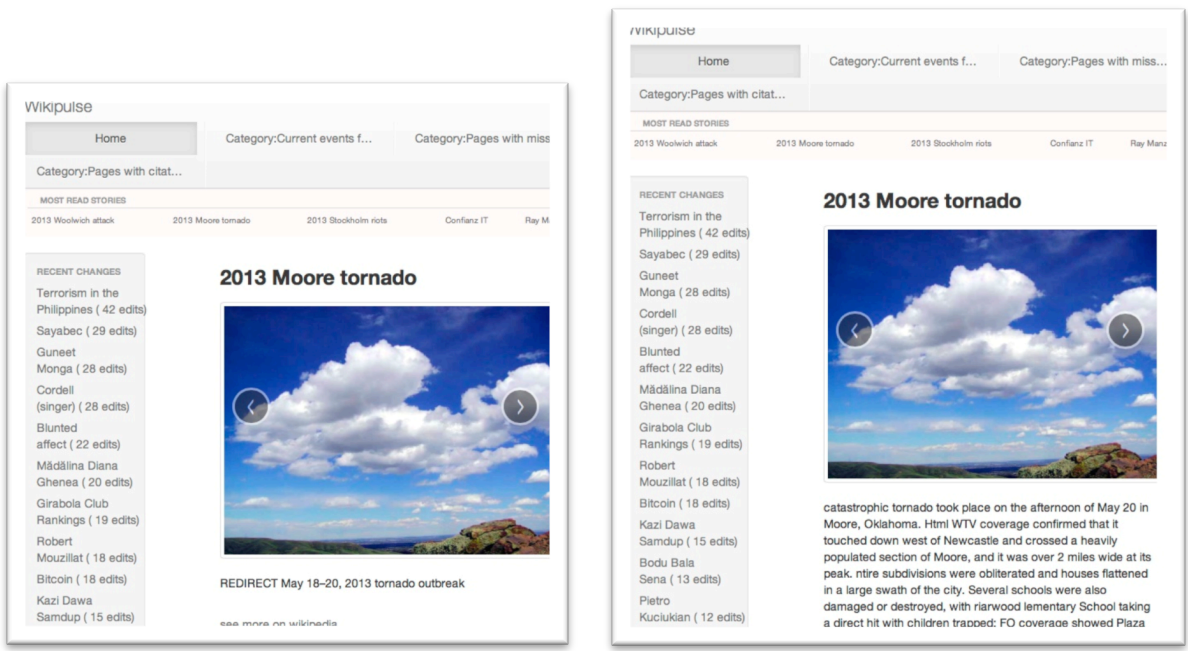

Figure 4. 2 Screen shots of the Wikipulse entry on the Oklahoma Moore tornado May 18, auto-generated on May 20, 2012 at 10:41PM (left) and seven minutes later, at May 20, 10:48PM (right)